\documentclass[preprint,showpacs,preprintnumbers,amsmath,amssymb,pre]{revtex4-1}

\usepackage{graphicx}
\usepackage{dcolumn}
\usepackage{bm}
\usepackage[latin1]{inputenc}
\usepackage{color}
\usepackage{amssymb}
\usepackage{amsmath}
\usepackage{mathrsfs}
\usepackage{multirow}
\usepackage{booktabs}
\usepackage[normalem]{ulem}

\graphicspath{{figures/},.}

\begin{document}
\preprint{APS/123-QED}

\title{Decelerated Spreading in Degree-Correlated Networks}

\author{Markus Schl\"apfer$^1$} 
\email{schlmark@mit.edu}
\author{Lubos Buzna$^{2}$}
\email{buzna@frdsa.uniza.sk}

\affiliation{$^1$Massachusetts Institute of Technology, Cambridge, MA 02139, USA\\
$^2$University of Zilina, SK-01026 Zilina, Slovakia}
\date{\today}
%
\begin{abstract}

While degree correlations are known to play a crucial role for spreading phenomena in networks, their impact on the propagation speed has hardly been understood. Here we investigate a tunable spreading model on scale-free networks and show that the propagation becomes slow in positively (negatively) correlated networks, if nodes with a high connectivity locally accelerate (decelerate) the propagation. Examining the efficient paths offers a coherent explanation for this result, while the  $k$-core decomposition reveals the dependence of the nodal spreading efficiency on the correlation. Our findings should open new pathways to delicately control real-world spreading processes.
\end{abstract}
\maketitle
Understanding the mechanisms of spreading phenomena is a need shared across many scientific disciplines, with examples as seemingly diverse as reaction diffusion processes, pandemics and cascading failures in electric power grids. Substantial new insights have recently been gained through the application of statistical physics to the study of large-scale networked systems, where extensive research has focused on two-point degree (or ``degree-degree'') correlations \cite{Boccaletti:2006,Barratt:2008,citeulike:4504463}. A network with a positive degree correlation is called assortative, and implies that nodes with a similarly small or large degree tend to be connected to each other \cite{PhysRevLett.89.208701}. If, by contrast, nodes tend to be connected to nodes with a considerably different degree, the network is called disassortative, referring to a negative correlation. Assortativity is typically found in social networks, and disassortativity is found in biological and technical networks \cite{PhysRevLett.89.208701}. The impact of correlations on spreading dynamics appears to be non-trivial \cite{Payne:2009} and has so far been discussed by modeling specific processes. Interestingly, assortativity seems to hinder disease spreading \cite{PhysRevE.68.035103} and information diffusion \cite{Karsai:2011}, while disassortativity has been suggested to prevent the propagation of perturbations in protein networks \cite{citeulike:99928} and to enhance the robustness of declining company networks \cite{saavedra08}. Nevertheless, regarding the impact on the spreading speed, a comprehensive picture is still lacking.

In this paper, we provide a first step toward filling this gap and show that many spreading models can be categorized into two types, for which either assortativity or disassortativity has a decelerating effect. By generalizing a spreading process as the cascading flipping of nodes from an initially \textit{inactive} (or \textit{susceptible}) to an \textit{active} (or \textit{infected}) state, the two types are then given by the neighborhood influence response function (NIRF) \cite{PhysRevE.77.046117}: in what we will call \mbox{\textit{type-I}} processes, nodes with a smaller degree are more likely to be activated than those with a larger degree, given that at least the same ratio of nearest neighbors is already in the active state (see illustration in Table \ref{tableA}). In \textit{type-II} processes, the activation probability is higher for nodes with a larger degree. These response rules are inherent to models for various phenomena. An example of type I is a model for a declining company network, where the probability for a company disappearing is inversely proportional to its degree \cite{saavedra08}. After losing the same ratio of connected firms a company with a large (initial) degree has still more connections and thus a lower probability of disappearing than a company with a small degree. An example of type II is the spreading model of epidemics on the air-transportation network \cite{Colizza:2006p19304}, since a highly linked city is more prone than a city with less links, given that the same ratio of connected cities has an equally infected population. Further examples are listed in Table \ref{tableA}.
\begin{table}[b]
\caption{\label{tableA}(color online). Exemplary models and their categorization based on the type of the NIRF. In the illustration, node $i$ changes its state with probability $P_i$ and node $j$ changes with $P_j$. The same ratio of nearest neighbors is already active [red (dark grey) color].}
%
\begin{tabular}{l|l }
\hline
\hline
\multirow{8}{*}{
\includegraphics[width=0.25\textwidth]{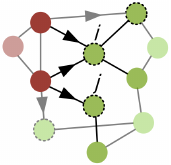} 
}& 
\multicolumn{1}{c}{\textit{Type I}: $P_i < P_j$}\\[0.1em]
& Declining company network \cite{saavedra08}\\[0.2em]
& Extinction of species\footnotemark[1] \cite{Staniczenko:2010} \\
\cline{2-2}
& \multicolumn{1}{c}{\textit{Type II}: $P_i > P_j$}\\[0.1em]

& Reaction-diffusion processes\footnotemark[2] \cite{Colizza:2007p3336}\\
& Global epidemics \cite{Colizza:2006p19304}\\ 
& Dissemination of information\footnotemark[3]  \cite{PhysRevE.69.055101} {}\\
& Cascading failures in power grids\footnotemark[4] \cite{PhysRevE.61.4877} \\
 \hline
 \hline
\end{tabular}

\footnotetext[1]{The probability of a species being removed is inversely proportional to its degree, reflecting the higher sensitivity of specialists to environmental stress.}
\footnotetext[2]{A node with a large degree has a higher probability of receiving active particles than a node with a small degree, if the same ratio of nearest neighbors has the same density of active particles.}
\footnotetext[3]{Large-degree nodes are likely to get the information at a lower ratio of nearest neighbors being already informed.}
\footnotetext[4]{If a certain ratio of nearest neighbors fails, a node with a larger degree has a higher probability of becoming overloaded.}
\end{table}

\begin{figure}[t!]
\includegraphics[width=0.74\textwidth]{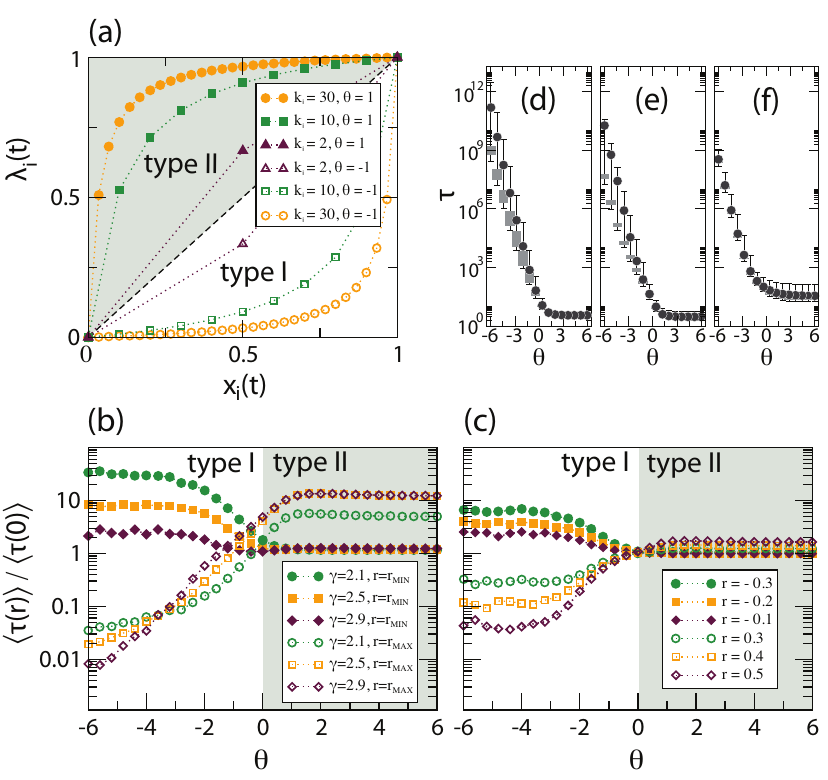}
\vspace{-0.3cm}
\caption{\label{fig:lifeExp1} (color online). (a) Activation rate vs the ratio of active nearest neighbors, for exemplary combinations of the degree and the response parameter. (b)(c) Normalized expected activation time in connected scale-free networks with $N=10^4$ and $k_\textrm{MIN}=2$ vs the response parameter, for different scale-free exponents and correlation levels: (b) $r_\textrm{MIN}=-0.55\pm0.02,-0.33\pm0.02,-0.21\pm0.02$ and $r_\textrm{MAX}=0.94\pm0.01,0.84\pm0.02,0.71\pm0.05$ for $\gamma=2.1,2.5,2.9$, respectively, and (c) $\gamma=2.5$ for all networks. The values are averages over a minimum of 10$^4$ simulation runs on 300 network realizations. (d)(e)(f) Expected activation time for $\gamma = 2.5$ and different correlation levels: (d) $r = r_\textrm{MIN}$, (e) $r = 0$, and (f) $r = r_\textrm{MAX}$. Dots represent $\left\langle \tau \right\rangle$, boxes range from lower to upper quartiles and whiskers from the 1st to the 99th percentile of the estimates from each simulation run.}
\end{figure}

In order to unravel the type-dependent effect of degree correlations on the propagation speed, we capture the spreading by the dynamic state variable $s_i(t) \in \{ 0, 1 \}$ assigned to each node $i$, with $s_i(t) = 1$ if the node is active and $s_i(t) = 0$ otherwise. To be completely general, we define $P_i(t)=\lambda_i(t) dt$ as the probability that in the interval $dt$ a node flips from $s_i(t) = 0$ to $s_i(t+dt) = 1$. Thus, $\lambda_i(t) = f_i(t)/[1-F_i(t)]$, with $F_i(t) = \int_0^t f_i(u)du$ being the activation time distribution function. The activation probability is required to be increasing with the ratio $x_i(t) = \sum_{j \in \mathcal{N}(i)} s_j(t) /k_i$ of activated neighboring nodes to the node degree $k_i$, with $\mathcal{N}(i)$ being the set of nearest neighbors of node $i$. In order to vary the type-dependent influence of $k_i$ and normalizing, so that $0 \leq \lambda_i(t) \leq 1$, let us define the activation rate $\lambda_{i}(x_{i}(t))$ here as

\begin{equation}\label{Eq:stress}
\displaystyle \lambda_i(x_i(t)) \equiv \frac{x_i(t) k_i^{\phantom{i}\theta}}{1+x_i(t) (k_i^{\phantom{i}\theta}-1)}.
\end{equation}
By tuning the response parameter $\theta$ we interpolate smoothly between the two spreading types, with type I for $\theta < 0$ and type II for $\theta > 0$ [Fig. \ref{fig:lifeExp1}(a)]. As we will show later, it is important to stress that Eq. (\ref{Eq:stress}) can be replaced by other functions that qualitatively reproduce the two response types.
Note that in the limit $\theta \gg 0$ we readily recover the susceptible-infected (SI) model for disease spreading, where $P_i(t)$ is the probability of acquiring the infection if at least one nearest neighbor is infected \cite{Anderson1992, Barratt:2008}. The proportional increase of $\lambda_{i}$, as $\theta=0$, corresponds to the linear NIRF of the Bass model for innovation diffusion \cite{1245934}. Further examples are binary threshold models for social contagion \cite{citeulike:1753065}; however, here the approximation is limited to the boundary cases with either very low or very high thresholds \cite{Note1}. 

The spreading is studied on degree-correlated scale-free networks with finite topological dimension. Their degree distribution follows a power law, $P(k) \propto k^{-\gamma}$, which is archetypical for many real networks with $2 < \gamma \leq 3$ \cite{Boccaletti:2006}. The global level of degree correlation is commonly quantified by the Pearson coefficient $r$, where $r = 0$ corresponds to an uncorrelated network and a positive ($r_+$) [negative ($r_-$)] value denotes positive (negative) correlation \cite{PhysRevLett.89.208701}. We first build ensembles of uncorrelated networks with $N$ nodes and scale-free exponent $\gamma$ according to the configuration model \cite{PhysRevE.71.027103}, restricting the degree $k_i$ of each node $i$ to $k_\textrm{MIN} \leq k_i \leq \sqrt{N}$ with $\sum_i k_i$ being even. Deploying these networks as null models, we subsequently apply the reshuffling method \cite{PhysRevE.70.066102,Menche:2010} to impose the desired correlation value in the bounded interval $[r_\textrm{MIN},r_\textrm{MAX}]$, while preserving the degree distribution \cite{Note3}. Simulations are initiated in a standard way, by setting the state variable of a randomly selected node to $s_i(0) = 1$, while all other nodes are inactive. We then monitor the activity increase throughout the whole network until $\sum_{i = 1}^N s_i(t) = N$. The spreading speed determines the expected time $\left\langle \tau \right\rangle$ until a randomly chosen node is activated: the slower the spreading is, the larger becomes its value. 

Estimating $\left\langle \tau \right\rangle$ by extensive Monte Carlo simulations shows evidence for disassortativity decelerating type-I processes and assortativity decelerating those of type II. This result is synthesized in Figs. \ref{fig:lifeExp1}(b)-1(c), comparing  $\left\langle \tau(r_+) \right\rangle$  for assortative with $\left\langle \tau(r_-)\right\rangle$ for disassortative networks, normalized by the corresponding values for the null models, $\left\langle \tau(0) \right\rangle$. For $\theta <0$ we find  $\left\langle \tau(r_+) \right\rangle < \left\langle \tau(0) \right\rangle < \left\langle \tau(r_-) \right\rangle$, being more pronounced for larger values of $|r|$. After marking a crossover in the intermediate range, $\left\langle \tau(r_+) \right\rangle \approx \left\langle \tau(0) \right\rangle \approx \left\langle \tau(r_-) \right\rangle$, the expected activation times become larger and thus the spreading slower in assortative networks when $\theta>0$, with $\left\langle \tau(r_-) \right\rangle \approx \left\langle  \tau(0) \right\rangle < \left\langle \tau(r_+) \right\rangle$. For $\theta \gg$ 0 the decelerating effect of the positive degree correlation persists, in agreement with \cite{PhysRevE.68.035103,Karsai:2011}. Owing to the very specific nature of the chosen NIRF [Eq. (\ref{Eq:stress})], $\left\langle \tau \right\rangle$ decreases dramatically when increasing $\theta$ [Figs. 1(d)-1(f)], and $\theta \ll 0$ implies a higher influence of the initial conditions, as reflected by a broader distribution of $\tau$ \cite{Note2}. The decelerating effect of the degree correlations does not, however, depend sensitively on the specific NIRF, as demonstrated next.

\begin{figure}[t!]
\includegraphics[width=0.74\textwidth]{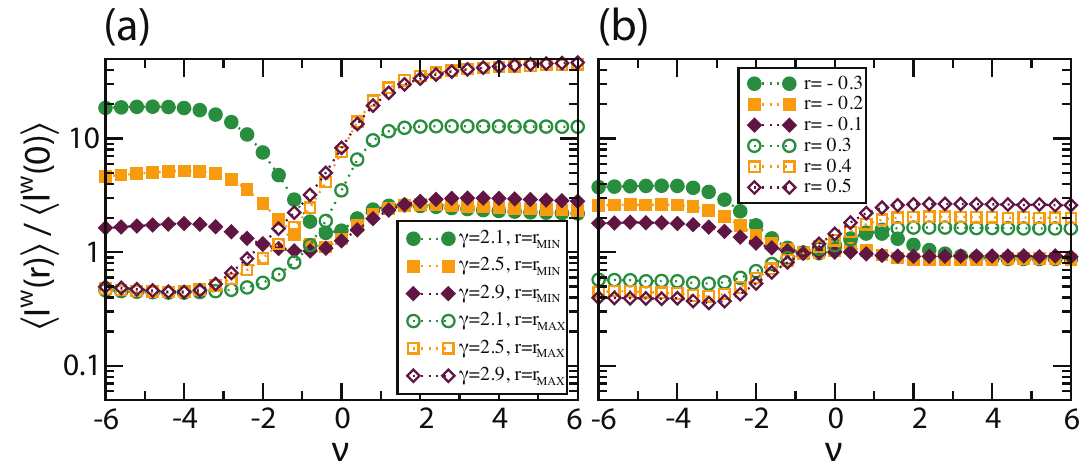}
\vspace{-0.3cm}
\caption{\label{fig:path}(color online). (a)(b) Average efficient path length of the same scale-free networks as in Figs. \ref{fig:lifeExp1}(b)-\ref{fig:lifeExp1}(c), normalized with the uncorrelated null model, vs the routing parameter: (a) correlation levels as in Fig. \ref{fig:lifeExp1}(b), and (b) $\gamma=2.5$ for all networks. The values are averages over 300 network realizations.}
\end{figure}

The correlation-dependent spreading speed can be rooted in the role of the nodes with a large degree and their location in the network. For $\theta < 0$ high-$k$ nodes are less sensitive to the states of the nearest neighbors than their sparsely connected counterparts, and we say that they act as propagation \textit{delayers}. In contrast, for $\theta > 0$, highly connected nodes are more affected by active nearest neighbors, and we say that they act as \textit{accelerators}. Thus, the propagation preferably bypasses through low-$k$ nodes in type-I processes and through high-$k$ nodes in type-II processes. In order to demonstrate this effect, we examined the `efficient paths' through which we expect the activation propagating most likely. Following \cite{yan:046108}, the length of a path $\mathcal{P}_{i,j}$, connecting node $i$ with node $j$ and containing the set of nodes $\mathcal{S_P}$, is given as 
\begin{equation}\label{Eq:path}
\displaystyle L^w(\mathcal{P}_{i,j};\nu) \equiv \sum_{\substack{\ell \in \mathcal{S_P}\\ \ell \neq j}} k_\ell^{-\nu},
\end{equation}
where $\nu$ is a parameter controlling the degree dependent path routing. The efficient path length is then the minimum value of $L^w(\mathcal{P}_{i,j};\nu)$ for all possible paths between nodes $i$ and $j$. Averaging over all pairs of nodes gives the average efficient path length $\left\langle l^{w}\right\rangle$, with $\nu = 0$ being the geodesic shortest path. As depicted in Fig. \ref{fig:path}, the values for $\left\langle l^{w}\right\rangle$ verify the numerical results of the spreading speed: for $\nu<0$ disassortative networks exhibit a larger value of $\left\langle l^{w}\right\rangle$, and for $\nu > 0$ the average efficient path is longer for assortative networks. 
Given this excellent agreement, the propagation indeed seems to follow the efficient paths, suggesting that $\left\langle l^{w}\right\rangle$ is a robust indicator for the impact of degree correlations on the spreading speed. More importantly, besides confirming the proposed model categorization, our results are not constrained on the particular NIRF [Eq. (\ref{Eq:stress})] and thus are applicable to a wider range of spreading processes. 

\begin{figure}[t!]
\includegraphics[width=0.74\textwidth]{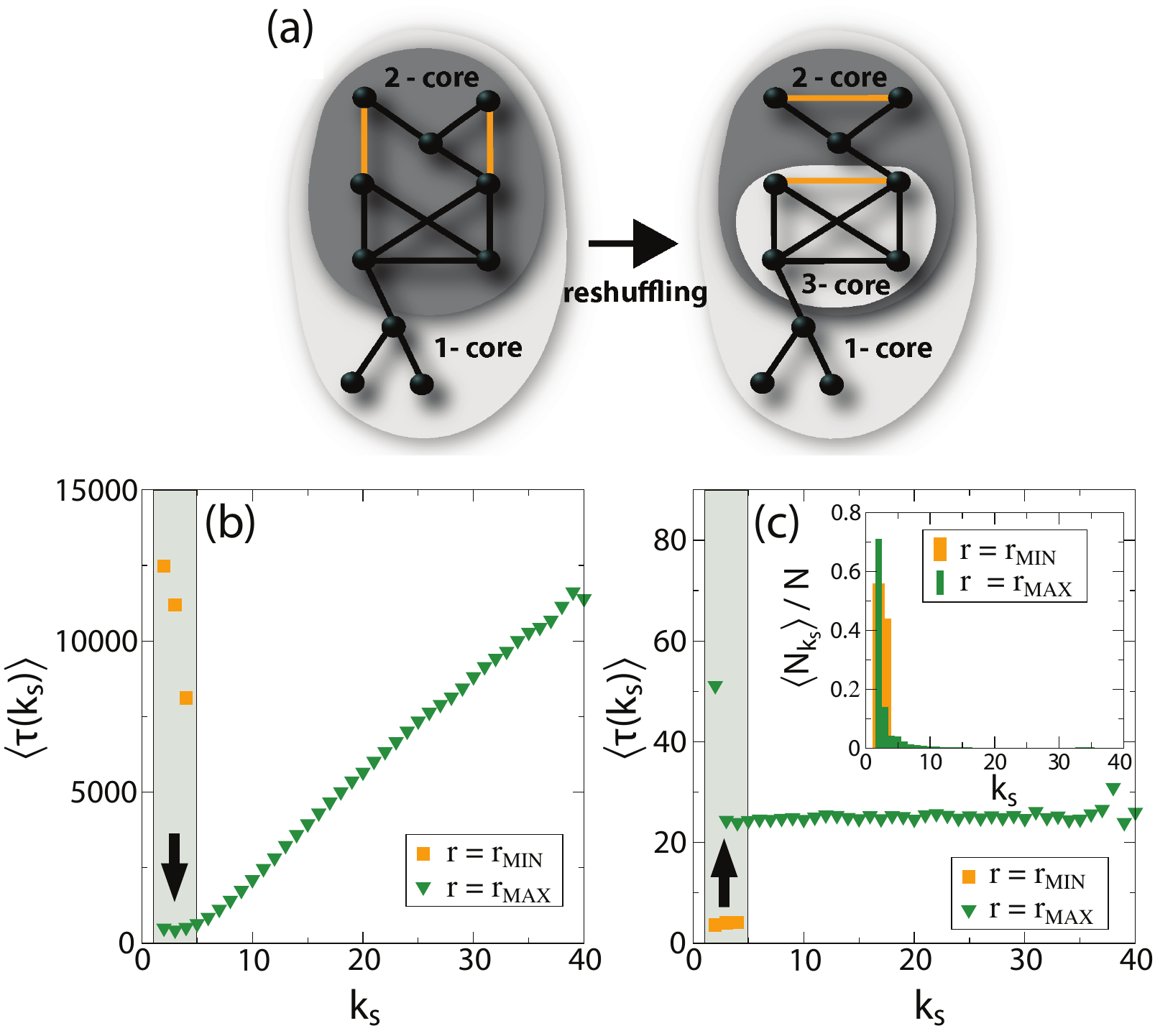}
\caption{\label{fig:cores}(color online). (a) Illustration of the $k$-core decomposition before (left) and after (right) reshuffling toward a higher positive correlation. Each shaded area corresponds to a different $k$-shell. (b)(c) Expected activation time of the $k$-shells in scale-free networks with $N = 10^4$, $\gamma = 2.5$ and $k_\textrm{MIN} = 2$ with minimum and maximum correlation: (b) $\theta = -2$, clustering the delayers within higher-order $k$-shells results in the bias of the fast propagation routes toward the network periphery; (c) $\theta = 2$, clustering the accelerators decelerates the propagation in the periphery. The arrows indicate the change of $\langle \tau(k_s) \rangle$ of the lower order $k$-shells when increasing the correlation level. The inset shows the relative size of the $k$-shells.}
\end{figure}

Seen from a different yet complementary angle, both accelerators and delayers become more efficient in disassortative networks. As the high-$k$ nodes are less topologically clustered than in assortative networks, the self-impeding overlap of their influenced areas is minimized. This phenomena has recently also been observed in real-world networks \cite{Kitsak:2010} and can be revealed by the $k$-core decomposition \cite{dorogovtsev-2006-96}. A $k$-core is the maximum subgraph with all nodes having minimum degree $k$, and a $k$-shell contains the fraction of nodes belonging to the $k$-core but not to the $(k+1)$-core, see Fig. \ref{fig:cores}(a). Clustering the delayers within higher-order $k$-shells \mbox{(type I, $r > 0$)} allows for a fast propagation in the lower-order $k$-shells [Fig. \ref{fig:cores}(b)], while clustering the accelerators (type II, $r > 0$) effectively decelerates the spreading within the lower-order $k$-shells [Fig. \ref{fig:cores}(c)].
In scale-free networks the majority of nodes remains in lower-order $k$-shells [Fig. \ref{fig:cores}(c), inset], so that these two opposite effects become directly reflected in $\left\langle \tau \right\rangle$.

\begin{figure}[t!]
\includegraphics[width=0.74\textwidth]{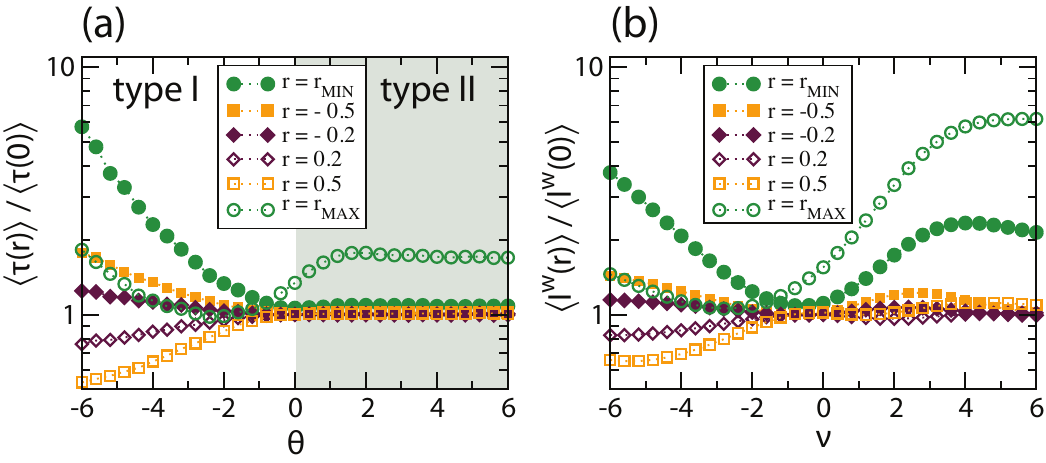}
\caption{\label{fig:poisson_nets}(color online).  (a) Normalized expected activation time in connected networks with a Poissonian degree distribution with $N=10^4$, average degree $\left\langle k \right\rangle = 5$, $k_\textrm{MIN} = 2$, $r_\textrm{MIN}=-0.94\pm0.004$ and $r_\textrm{MAX}=0.98\pm0.002$. The values are averages over $10^4$ simulation runs on 300 network realizations. (b) Corresponding average efficient path length.}
\end{figure}

The decelerating effect of degree correlations thus relies on a rich network topology, being a very natural feature of real-world networks \cite{Kitsak:2010}. In order to highlight the importance of large degree fluctuations, Fig. \ref{fig:poisson_nets} depicts the behavior of $\left\langle \tau \right\rangle$ for networks with a Poissonian degree distribution. Indeed, for $\theta < 0$ and $r > 0$ the activation times first decrease with growing $r$, as observed in scale-free networks, but increase again for $r = r_\textrm{MAX}$ [Fig. \ref{fig:poisson_nets}(a)], in agreement with \cite{Payne:2009}. This non-monotonous behavior is again confirmed by the efficient paths [Fig. \ref{fig:poisson_nets}(b)].
\newpage

In summary, we have drawn a global picture on how the spreading speed is jointly determined by the NIRF and the degree correlations in the underlying network. By introducing a tunable model allowing us to interpolate between two fundamental spreading types, we were able to reveal that the propagation becomes slow in assortative networks, if high-$k$ nodes locally act as accelerators. Conversely, the propagation becomes slow in disassortative networks, if the high-$k$ nodes act as delayers. Exploiting this opposite yet dramatic effect should provide efficient strategies to delicately control many real-world spreading processes, so as to impede epidemic diseases or to accelerate the diffusion of information.

The authors thank Adilson E. Motter and Albert D\'iaz-Guilera for helpful discussions and the anonymous referees for their valuable suggestions. M.S. acknowledges Swisselectric Research for co-funding the present work. L.B. gratefully acknowledges partial financial support from the project VEGA 1/0361/10.

%

\newpage
\renewcommand{\thefigure}{S\arabic{figure}}
\setcounter{figure}{0}

\renewcommand{\theequation}{S\arabic{equation}}
\begin{center}
\vspace{12cm}
\LARGE
{Supplemental material}
\end{center}
\normalsize

This supplemental material discusses the decelerating effect of two-point degree correlations on spreading processes described by Watts' threshold model of social contagion, and extends the study to scale-free networks with copula-based correlation structures.
\vspace{1cm}

\section{Threshold model}

\subsection{Neighborhood influence response function}

Applying the definitions given in the main text to Watt's threshold model \cite{SWatts:2002}, each node $i$ of a network can be in one of two possible states $s_i(t) \in \{ 0, 1 \}$, with $s_i(t) = 0$ if the node is \textit{inactive} (or \textit{susceptible}) and $s_i(t) = 1$ if the node is \textit{active} (or \textit{infected}). Once a node is active, it cannot deactivate. The nodes follow a threshold-based neighborhood influence response function (NIRF), where the activation probability $P_i(t)$ is a function of the ratio  $x_i(t) = \sum_{j \in\mathcal{N}(i)} s_j(t)/k_i$ of active nearest neighbors to the degree $k_i$, with $\mathcal{N}(i)$ being the set of nearest neighbors. Specifically, if $\phi$ denotes the identical threshold for all nodes, then the NIRF is given as
\begin{equation} 
\label{eq:threshold}
P_i(t)  =\begin{cases} 1 & \text{if } x_i(t) \geq \phi,\\ 0 & \textrm{otherwise}.
\end{cases} \end{equation}

For either very low or very high values of $\phi$, the threshold model is directly related to the specific NIRF used in the main text [Eq. (1)]. Low thresholds (i.e., $\phi \approx 0$) can be approximated by $\theta \gg 0$, as 

\begin{equation} \label{eq:approx}P_i(t)=\lim_{\theta \to +\infty} \frac{x_i(t) k_i^\theta}{1+x_i(t)(k_i^\theta-1)} \cdot dt =
\begin{cases} 0 & \text{if } x_i(t) = 0,\\ 1\cdot dt & \text{if } 0 < x_i(t)\leq 1.
\end{cases} \end{equation}
Similarly, high thresholds (i.e., $\phi \approx 1$) can be approximated by $\theta \ll 0$, as
\begin{equation} P_i(t)=\lim_{\theta \to -\infty} \frac{x_i(t) k_i^\theta}{1+x_i(t)(k_i^\theta-1)} \cdot dt  =
\begin{cases} 0 & \text{if } 0 \leq x_i(t) < 1,\\1 \cdot dt & \text{if } x_i(t) = 1.
\end{cases} \end{equation}

\subsection{Spreading speed}

\begin{figure}
\centering
\includegraphics[width=15cm]{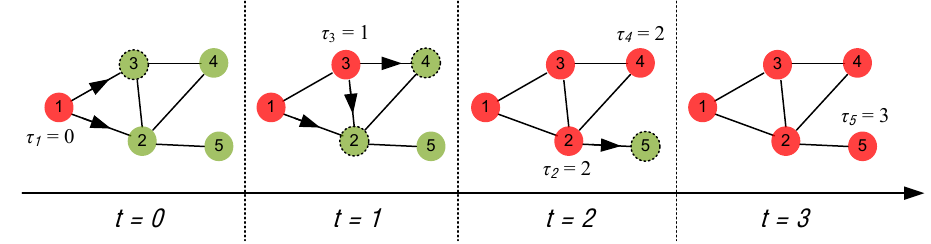}
\caption{(color online). Schematic illustrating the activation time as measured for Watts' threshold model ($\phi=0.3$). At time $t=0$ node 1 is switched to the active state and triggers the subsequent cascade.}
\label{schematic}
\end{figure}

Similar studies so far have analyzed the effect of degree correlations on the frequency and size of cascades triggered by activating a single node (e.g., \cite{SDodds:2009,SPayne:2009}). The approach here, however, differs from these studies as \textit{i)} we are assessing the \textit{speed} of the propagation and \textit{ii)} in order to ensure the analogy with the main text we shall focus on \textit{full} cascades, where all nodes eventually are active. At time $t=0$ all nodes are inactive and the cascade is triggered by switching a randomly selected node to the active state. The activation evolves throughout the complete network at successive time steps with all nodes updating their states according to the threshold rule [Eq. (\ref{eq:threshold})]. We then estimate the expected activation time $\langle \tau \rangle$ by averaging over the individual activation times $\tau_i$, being measured according to the schematic shown in Fig. \ref{schematic}: a low (high) value of $\langle \tau \rangle$ indicates fast (slow) spreading. Additionally, we evaluate the frequency $F_{fc}$ of full cascades in order to assess their relevance.

\subsection{Numerical experiments}

\begin{figure}[t!]
\centering
\includegraphics[width=13cm]{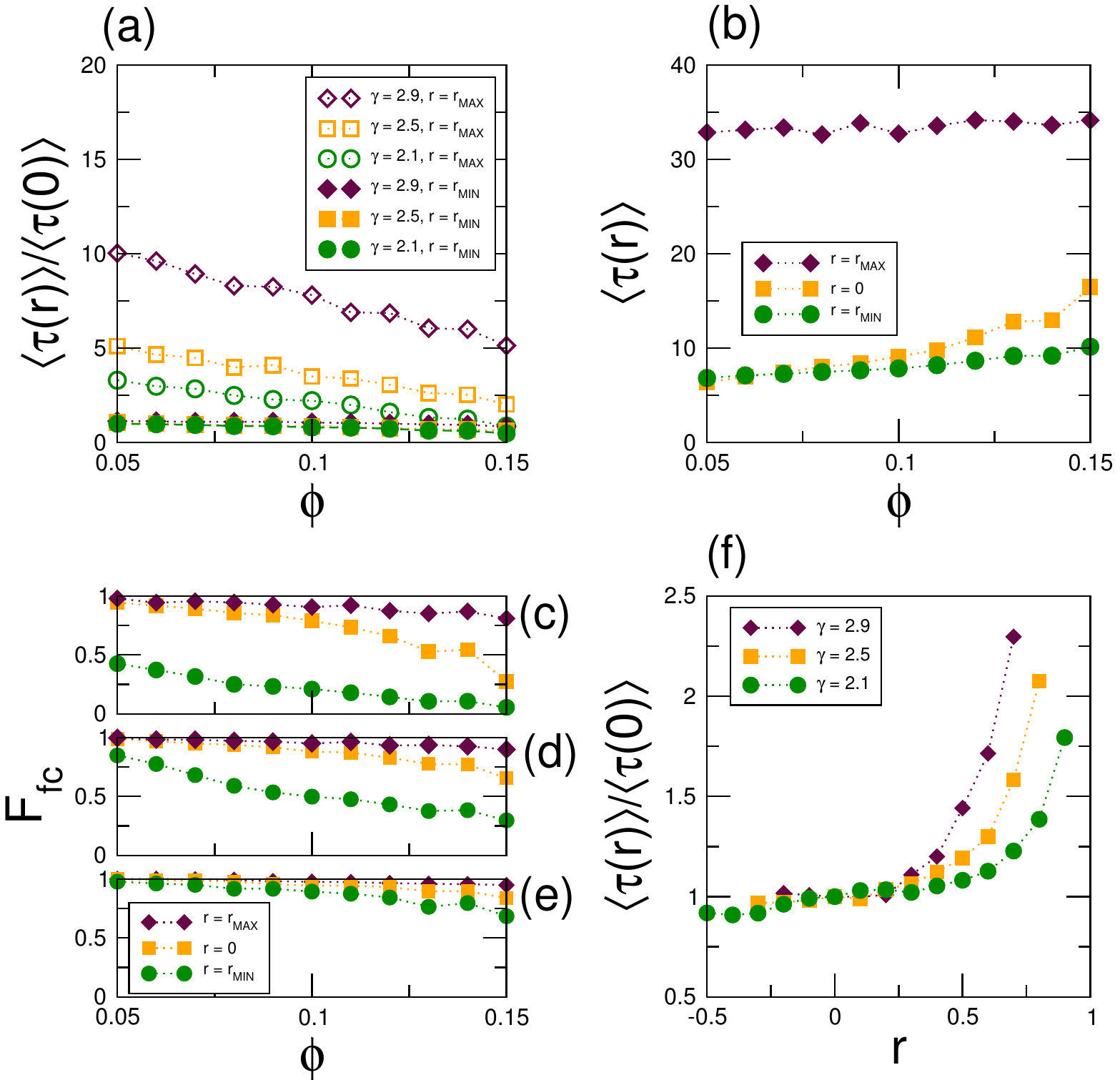}
\caption{(color online). (a) Normalized expected activation time in connected scale-free networks with $N=4000$ and $k_\textrm{MIN}=2$ versus the threshold for different values of the characteristic exponent and different correlation levels: $r_\textrm{MIN}=-0.61\pm0.02,-0.40\pm0.02,-0.26\pm0.03$ and $r_\textrm{MAX}=0.95\pm0.01,0.86\pm0.03,0.78\pm0.05$ for $\gamma = 2.1, 2.5, 2.9$, respectively. (b) Absolute value of the expected activation time for $\gamma = 2.5$ and different correlation levels. (c)-(e) Frequency of full cascades for the same networks as in panel (a): (c) $\gamma = 2.1$, (d) $\gamma = 2.5$ and (e) $\gamma = 2.9$. \mbox{(f) Normalized} expected activation time versus the correlation level for $\phi=0.05$. The results are based on a minimum of 400 simulation runs performed on 50 network realizations.}
\label{fig:fig2}
\end{figure}

Analogously to the experiments presented in the main text, the spreading speed is studied by extensive numerical simulations. All experiments are performed on scale-free networks with $N = 4000$ and $k_\textrm{MIN} = 2$, whereas the degree correlations are altered within the bounded interval $[r_\textrm{MIN}, r_\textrm{MAX}]$. As the frequent occurrence of full cascades requires sufficiently low thresholds, we consider values of $\phi$ within the interval $[0.05, 0.15]$ in increments of 0.01. The numerical results are summarized in Fig. \ref{fig:fig2}.
In alignment with the arguments given in the main text, the propagation is slower in assortative networks [Figs. \ref{fig:fig2}(a)-\ref{fig:fig2}(b)]. For small values of $\phi$ the threshold model closely resembles spreading processes of type II [see \mbox{Eq. (\ref{eq:approx}),} $\theta \gg 0$], whereas already a small fraction of activated nearest neighbors turns a node to the active state. Thus, following the explanations given in the main text, high-$k$ nodes act as propagation accelerators. In positively correlated networks these nodes are clustered in the core of the network and thus become less efficient, consequently suppressing a fast propagation. Vice versa, the propagation speed is significantly higher if accelerators are more uniformly distributed across the network, which applies to uncorrelated or disassortative networks.  As could be expected, increasing $\phi$ reduces the frequency of full cascades [Figs.~\ref{fig:fig2}(c)-\ref{fig:fig2}(e)]. Being again consistent with the findings presented in the main text, the deceleration significantly increases with the correlation level [Fig.~\ref{fig:fig2}(f)]. For $r<0$ the propagation speed is close to the uncorrelated case, but with increasing value of $r$ the propagation slows down.

Taken together, Watt's threshold model with low values of $\phi$ is a type-II process according to the classification of spreading models suggested in the main text. As numerically confirmed, the propagation consequently is slower in positively correlated networks, where the high-$k$ nodes (acting as accelerators) are clustered in the network core.

\section{Networks with copula-based correlation structures}

\begin{figure}[t!]
\centering
\includegraphics[width=15cm]{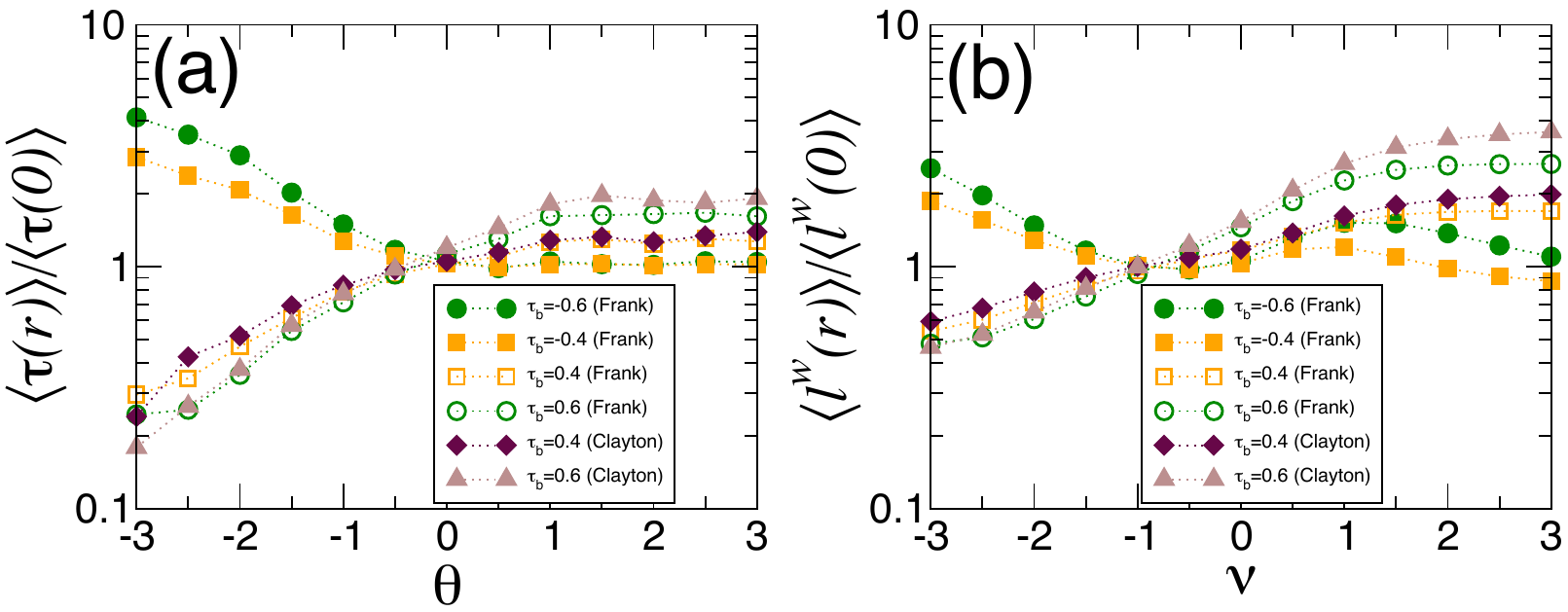}
\caption{(color online). (a) Normalized expected activation time in connected scale-free networks with $N=4000$, $\gamma=2.5$, $k_\textrm{MIN}=2$, and different copula functions versus the response parameter. The parameter $\tau_b$ indicates Kendall's rank correlation coefficient for the underlying probability matrix. The correlation levels of the realized networks are quantified by an average Pearson coefficient of $\langle r \rangle$=0.35,0.52,-0.29,-0.36 for the Frank copulas with $\tau_b$=0.4.0.6,-0.4,-0.6, and $\langle r \rangle$=0.24,0.4 for the Clayton copulas with $\tau_b$=0.4,0.6, respectively. The results are based on a minimum of 500 simulation runs performed on 100 network realizations. (b) Corresponding average efficient path length.}
\label{Fig:SI3}
\end{figure}

The commonly applied reshuffling method according to \cite{SXulvi-Brunet:2004,SMenche:2010} is used in the main text to impose a desired level of two-point degree correlation on networks. This straightforward algorithm has limited capabilities to fully control the overall correlation \textit{structure}, so that the latter may differ for two networks with equal value of the correlation measure $r$. In order to assess whether the decelerating effect as described in the main text also holds for different correlation structures, we applied the copula-based network generation algorithm as introduced in \cite{SRaschke:2010}. Copulas enable the realization of random network ensembles based on a probability matrix with an \textit{a priori} determined correlation structure. The correlation level of the probability matrix is thereby quantified by Kendall's rank correlation coefficient \mbox{$\tau_b$ \cite{SRaschke:2010b}}. The procedural details are given in \cite{SRaschke:2010}. 
For our study, we applied the tunable spreading process [Eq. (1) in the main text]  to scale-free networks based on probability matrices which have been generated by Frank and Clayton copulas \cite{SMari:2001}. The degree $k_i$ of each node $i$ is thereby again restricted to $k_\textrm{MIN} \leq k_i \leq \sqrt{N}$. Figure \ref{Fig:SI3}(a) shows the resulting estimates of the expected nodal activation times, $\langle \tau \rangle$, for the two selected copulas and different values of $\tau_b$. 

The expected activation times are not sensitive to altering the correlation structure by applying different copula functions, while preserving the correlation level $\tau_b$ of the underlying probability matrix and thus of the realized networks. Furthermore, the impact of different correlation levels on the spreading speed is fully consistent with networks generated by the reshuffling algorithm [see Figs. 1(b)-1(c) in the main text]. Thereby, the simulation results are again in excellent agreement with the efficient paths, see \mbox{Fig. \ref{Fig:SI3}(b)}. Hence, applying scale-free networks derived from copula functions further confirms the decelerating effect of two-point degree correlations on a wider class of networks.

\end{document}